\documentclass[a4paper]{article}
\usepackage{amsmath,amsthm,mathrsfs}
\usepackage{graphicx}
\usepackage[dvipsnames]{xcolor}
\usepackage{INTERSPEECH2020}
\usepackage{caption}
\usepackage[hyphens]{url}
\usepackage{subcaption}
\usepackage{hhline}
\usepackage{makecell}
\usepackage{booktabs}
\usepackage{pgfplots}
\pgfplotsset{compat=1.9}
\usepackage{float}
\usepackage[final]{changes}
    
\usepackage{balance}

\newcommand{\Cllrmin}{C_{\text{llr}}^{\text{min}}}
\newcommand{\Dece}{D_{\text{ECE}}}

\title{Adversarial Disentanglement of Speaker Representation \\ for Attribute-Driven Privacy Preservation}
\name{Paul-Gauthier Noé$^1$, Mohammad Mohammadamini$^1$, Driss Matrouf$^1$, \\Titouan Parcollet$^{1,2}$, Andreas Nautsch$^3$, Jean-François Bonastre$^1$}

\address{
  $^1$Laboratoire Informatique d'Avignon (LIA), Avignon Université, France\\
  $^2$University of Cambridge, United-Kingdom\\
  $^3$Digital Security Department, EURECOM, France}
\email{paul-gauthier.noe@univ-avignon.fr}

\begin{document}

\maketitle  
\begin{abstract}
In speech technologies, speaker's voice representation is used in many applications such as speech recognition, voice conversion, speech synthesis and, obviously, user authentication. Modern vocal representations of the speaker are based on neural embeddings. In addition to the targeted information, these representations usually contain sensitive information about the speaker, like the age, sex, physical state, education level or ethnicity. In order to allow the user to choose which information to protect, we introduce in this paper the concept of \emph{attribute-driven privacy preservation} in speaker voice representation. It allows a person to hide one or more personal aspects to  
a potential malicious interceptor and to the application provider.
As a first 
solution to this concept, we propose to use an adversarial autoencoding method that disentangles in the voice representation a given speaker attribute thus allowing its concealment. We focus here on the sex attribute for an Automatic Speaker Verification (ASV) task. Experiments carried out using the VoxCeleb datasets have shown that the proposed method enables the concealment of this attribute while preserving ASV ability.
\end{abstract}


\section{Introduction}

Data protection regulation such as the European \emph{General Data Protection Regulation} (GDPR) and the privacy concern in speech technologies~\cite{Nautsch2019gdpr} call for more control on the personal information an user discloses while using such technologies. This paper presents and extends the \emph{attribute-driven privacy preservation} introduced in~\cite{noe2020adversarial}. The idea is to let the user decide 
what personal information they agree to disclose in their voice data while using a given voice service. 
In this work, we focus our efforts on the widely used x-vector~\cite{snyder2018x} speaker's voice representation. Despite its effectiveness for many applications, this representation is known to contain sensitive information~\cite{raj2019probing}. To answer this privacy issue, most of the speech privacy preservation systems impact the full speaker's representation~\cite{pathak2012,NAUTSCH2019441,fuming2019,srivastava2019evaluating, Tomashenko2020} while in some use cases it is necessary and/or sufficient to hide only one or a few personal attributes in order to maintain the performance of the vocal system or to preserve the vocal richness as much as possible.
To overcome this drawback, we propose to consider the disentanglement learning~\cite{bengio2013representation} to facilitate the control over information in speaker's voice representations.

Various works have been proposed on linguistic/non-linguistic~\cite{acvae,Hsu2017,pmlr-v80-yingzhen18a,fdmm} and speaker/noise information disentanglement~\cite{hsu2019disentangling}. In~\cite{Srivastava2019} an adversarial approach is used 
to hide the speaker's identity in Automatic Speech Recognition (ASR) representations.
Most of the investigations on disentanglement methods are conducted directly on acoustic feature sequences, and few works have been proposed to operate on speaker representations. In~\cite{williams2019disentangling}, an autoencoder is applied on x-vectors to disentangle the speaking style and the identity of the speaker within two subspaces. More recently, in~\cite{Peri2020Robustspeakerrecognitionusing}, an unsupervised disentanglement method have been used over x-vectors to separate the speaker-discriminative information and the remaining noises for robust speaker recognition. However, none of these works considered to disentangle soft-biometric attributes directly from speaker representations. In~\cite{nelus2018}, an adversarial approach is proposed for the extraction of gender-discriminative features with low speaker-discriminative information. Whereas this approach aims to enable the detection of a speaker attribute (the sex\footnote{Throughout this paper, the term \emph{sex} refers to the biological differences between female and male~\cite{PrinceVirginia2005SvG}.}) while avoiding speaker identification, our work allows the contrary: enabling speaker verification while avoiding the detection of a specific speaker attribute.

Indeed, as a first example and solution to the attribute-driven privacy preservation idea, this paper presents a method to disentangle and hide the sex attribute in x-vector embeddings using an adversarial autoencoding approach. Our goal is to allow the user to hide this specific personal attribute in the x-vector while preserving the remaining information in order to perform authentication-by-voice. 
While we focus here on ASV, this approch can be experimented on any application that is based on such speaker representation from disease detection~\cite{Botelho2020PathologicalSD} to voice conversion and anonymisation~\cite{fuming2019} for instance. The effectiveness of the method is assessed on sex concealment evaluation and speaker verification tasks using the VoxCeleb corpora~\cite{voxceleb,voxceleb2}. 

Section~\ref{sec:task} explains the attribute-driven privacy preservation concept; Section~\ref{sec:adv_dis_autoencod} presents the adversarial disentangling autoencoder\footnote{Code is available at 
\label{git}\url{https://github.com/LIAvignon/adversarial-disentangling-autoencoder-for-spk-representation}} for hiding a binary attribute in x-vectors; Section~\ref{sec:global_exps} presents the experimental setup and the evaluation metrics
; The results on the VoxCeleb corpora are then presented and discussed in Section~\ref{sec:results}; Finally, Section~\ref{sec:conclusion} concludes and presents potential future works.



\section{Attribute-driven privacy preservation}
\label{sec:task}
The idea behind the attribute-driven privacy preservation is to not disclose one or a few personal information while enabling a desired task such as automatic speech recognition or automatic speaker verification for instance. This idea has been independently presented in~\cite{aloufi2020privacy} and is referred as \emph{user-configurable privacy}.
As a first toy example, this paper focuses on the sex attribute and automatic speaker verification.

\begin{figure}[!ht]
    \centering
    \includegraphics[width=7cm]{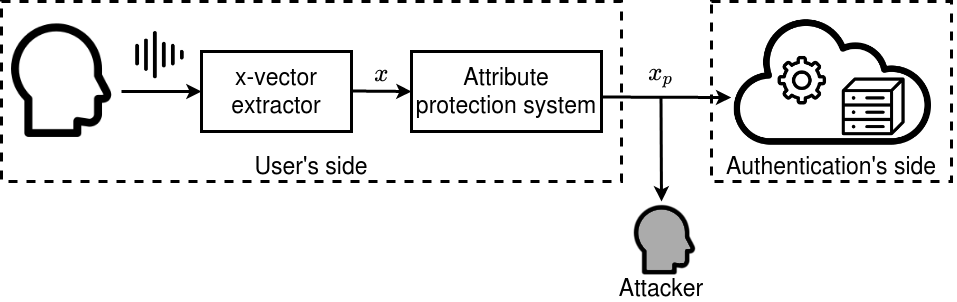}
    \caption{Example of an attribute-driven privacy preservation use case. The released data $x_p$ are processed by the authentication side but may also be intercepted by an unexpected attacker. Therefore, the user employs an attribute protection system in order to not disclose some aspects s/he decided to keep secret.}
    \label{fig:adpp_asv}
\end{figure}

Figure~\ref{fig:adpp_asv} illustrates the situation where an user conceals some information in its voice representation in order to avoid the authentication side and an attacker to infer these personal information. Indeed, they could try to detect for example the sex of the user using scores obtained from a sex classifier that operates on x-vectors. The role of the protection system is thus to make these scores unexploitable.

\section{Adversarial disentangling autoencoder and attribute protection}
%
\label{sec:adv_dis_autoencod}

This section presents a first solution to conceal the sex of the speaker in its x-vector representations.

\subsection{Adversarial disentangling autoencoder}

The proposed model, similarly to the one used in~\cite{lample2017fader} for image processing, disentangles the sex information in x-vector speaker representations. It has four components: a pre-trained sex classifier, an encoder, a decoder and an adversarial sex classifier.\\
\textbf{Notation:} let $D=\{(x_{1},y_{1},\tilde{y}_{1}),...(x_{m},y_{m},\tilde{y}_{m})\}$ be a set of x-vectors with their binary sex label (0 for male, 1 for female) and as a soft label, the posterior probability $\tilde{y}=P(F|x)$ where $F$ is the proposition: \emph{"the speech segment has been uttered by a female speaker"}.\\
\textbf{The pre-trained sex classifier} is used as a feature extractor that predicts the posteriors $\tilde{y}$ as soft labels. Soft labelling is used instead of hard labelling to make the \emph{sex} variable not strongly binary and to allow values between 0 and~1. Moreover, having probabilistic interpretation would be consistent with the Bayesian decision framework. As shown in Figure~\ref{fig:empcalplot}, the scores might not correspond to proper posterior probabilities. Therefore, a calibration step~\cite{pav_brummer} is added in order to produce oracle calibrated scores.
\begin{figure}[h]
    \centering
    \subfloat[Without calibration]{
\begin{tikzpicture}[font=\scriptsize]
\definecolor{color0}{rgb}{0.12156862745098,0.466666666666667,0.705882352941177}

\begin{axis}[
height=0.21\textwidth,
legend columns=1,
legend pos=outer north east,
tick align=outside,
tick pos=left,
width=0.23\textwidth,
x grid style={white!69.0196078431373!black},
xmin=0, xmax=1,
xtick style={color=black},
xtick={0,0.2,0.4,0.6,0.8,1},
xticklabels={$0$,$0.2$,$0.4$,$0.6$,$0.8$,$1$},
xlabel={Empirical posterior},
y grid style={white!69.0196078431373!black},
ymin=0, ymax=1,
ytick style={color=black},
ytick={0,0.2,0.4,0.6,0.8,1},
yticklabels={$0$,$20$,$40$,$60$,$80$,$100$},
ylabel = {Proportion \\ of female [\%]},
ylabel style={align=center} 
]
\draw[fill=color0,draw opacity=0] (axis cs:0.095,0) rectangle (axis cs:0.11,0);
\draw[fill=color0,draw opacity=0] (axis cs:0.115,0) rectangle (axis cs:0.13,0.00740740740740741);
\draw[fill=color0,draw opacity=0] (axis cs:0.135,0) rectangle (axis cs:0.15,0.00403877221324717);
\draw[fill=color0,draw opacity=0] (axis cs:0.155,0) rectangle (axis cs:0.17,0.00620067643742954);
\draw[fill=color0,draw opacity=0] (axis cs:0.175,0) rectangle (axis cs:0.19,0.00786051164784908);
\draw[fill=color0,draw opacity=0] (axis cs:0.195,0) rectangle (axis cs:0.21,0.00801915475828519);
\draw[fill=color0,draw opacity=0] (axis cs:0.215,0) rectangle (axis cs:0.23,0.00888481195125147);
\draw[fill=color0,draw opacity=0] (axis cs:0.235,0) rectangle (axis cs:0.25,0.010112714631834);
\draw[fill=color0,draw opacity=0] (axis cs:0.255,0) rectangle (axis cs:0.27,0.0130302081702927);
\draw[fill=color0,draw opacity=0] (axis cs:0.275,0) rectangle (axis cs:0.29,0.0154402102496715);
\draw[fill=color0,draw opacity=0] (axis cs:0.295,0) rectangle (axis cs:0.31,0.0191757430311981);
\draw[fill=color0,draw opacity=0] (axis cs:0.315,0) rectangle (axis cs:0.33,0.0266070367120496);
\draw[fill=color0,draw opacity=0] (axis cs:0.335,0) rectangle (axis cs:0.35,0.0384877751050156);
\draw[fill=color0,draw opacity=0] (axis cs:0.355,0) rectangle (axis cs:0.37,0.06281234688878);
\draw[fill=color0,draw opacity=0] (axis cs:0.375,0) rectangle (axis cs:0.39,0.0953634338935007);
\draw[fill=color0,draw opacity=0] (axis cs:0.395,0) rectangle (axis cs:0.41,0.154983983418127);
\draw[fill=color0,draw opacity=0] (axis cs:0.415,0) rectangle (axis cs:0.43,0.247124968055201);
\draw[fill=color0,draw opacity=0] (axis cs:0.435,0) rectangle (axis cs:0.45,0.375690607734807);
\draw[fill=color0,draw opacity=0] (axis cs:0.455,0) rectangle (axis cs:0.47,0.503572771718691);
\draw[fill=color0,draw opacity=0] (axis cs:0.475,0) rectangle (axis cs:0.49,0.63001638001638);
\draw[fill=color0,draw opacity=0] (axis cs:0.495,0) rectangle (axis cs:0.51,0.755182421227197);
\draw[fill=color0,draw opacity=0] (axis cs:0.515,0) rectangle (axis cs:0.53,0.828665250144872);
\draw[fill=color0,draw opacity=0] (axis cs:0.535,0) rectangle (axis cs:0.55,0.881623641304348);
\draw[fill=color0,draw opacity=0] (axis cs:0.555,0) rectangle (axis cs:0.57,0.915716856628674);
\draw[fill=color0,draw opacity=0] (axis cs:0.575,0) rectangle (axis cs:0.59,0.938073106393669);
\draw[fill=color0,draw opacity=0] (axis cs:0.595,0) rectangle (axis cs:0.61,0.950341207349081);
\draw[fill=color0,draw opacity=0] (axis cs:0.615,0) rectangle (axis cs:0.63,0.960838807478524);
\draw[fill=color0,draw opacity=0] (axis cs:0.635,0) rectangle (axis cs:0.65,0.970953796271278);
\draw[fill=color0,draw opacity=0] (axis cs:0.655,0) rectangle (axis cs:0.67,0.978139485882849);
\draw[fill=color0,draw opacity=0] (axis cs:0.675,0) rectangle (axis cs:0.69,0.983110051779403);
\draw[fill=color0,draw opacity=0] (axis cs:0.695,0) rectangle (axis cs:0.71,0.987450029490792);
\draw[fill=color0,draw opacity=0] (axis cs:0.715,0) rectangle (axis cs:0.73,0.991779105879897);
\draw[fill=color0,draw opacity=0] (axis cs:0.735,0) rectangle (axis cs:0.75,0.993301393917059);
\draw[fill=color0,draw opacity=0] (axis cs:0.755,0) rectangle (axis cs:0.77,0.994367366299063);
\draw[fill=color0,draw opacity=0] (axis cs:0.775,0) rectangle (axis cs:0.79,0.99446877634528);
\draw[fill=color0,draw opacity=0] (axis cs:0.795,0) rectangle (axis cs:0.81,0.995632114875379);
\draw[fill=color0,draw opacity=0] (axis cs:0.815,0) rectangle (axis cs:0.83,0.996051919956733);
\draw[fill=color0,draw opacity=0] (axis cs:0.835,0) rectangle (axis cs:0.85,0.9961262831687);
\draw[fill=color0,draw opacity=0] (axis cs:0.855,0) rectangle (axis cs:0.87,0.998168498168498);
\draw[fill=color0,draw opacity=0] (axis cs:0.875,0) rectangle (axis cs:0.89,1);
\addplot [thick, color0, dashed]
table {%
0 0
1 1
};
\end{axis}

\end{tikzpicture}
    }
    \hfill
    \subfloat[With oracle calibration]{
\begin{tikzpicture}[font=\scriptsize]
\definecolor{color0}{rgb}{0.12156862745098,0.466666666666667,0.705882352941177}

\begin{axis}[
height=0.21\textwidth,
legend columns=1,
legend pos=outer north east,
tick align=outside,
tick pos=left,
width=0.23\textwidth,
x grid style={white!69.0196078431373!black},
xmin=0, xmax=1,
xtick style={color=black},
xtick={0,0.2,0.4,0.6,0.8,1},
xticklabels={$0$,$0.2$,$0.4$,$0.6$,$0.8$,$1$},
xlabel={Empirical posterior},
y grid style={white!69.0196078431373!black},
ymin=0, ymax=1,
ytick style={color=black},
ytick={0,0.2,0.4,0.6,0.8,1},
yticklabels={$0$,$20$,$40$,$60$,$80$,$100$},
ylabel = {},
ylabel style={align=center} 
]
\draw[fill=color0,draw opacity=0] (axis cs:-0.005,0) rectangle (axis cs:0.01,0.011874990411022);
\draw[fill=color0,draw opacity=0] (axis cs:0.015,0) rectangle (axis cs:0.03,0.0282687506171446);
\draw[fill=color0,draw opacity=0] (axis cs:0.035,0) rectangle (axis cs:0.05,0.0485603215391711);
\draw[fill=color0,draw opacity=0] (axis cs:0.055,0) rectangle (axis cs:0.07,0.0696197912865796);
\draw[fill=color0,draw opacity=0] (axis cs:0.075,0) rectangle (axis cs:0.09,0.0924879681941829);
\draw[fill=color0,draw opacity=0] (axis cs:0.095,0) rectangle (axis cs:0.11,0.105776839883936);
\draw[fill=color0,draw opacity=0] (axis cs:0.115,0) rectangle (axis cs:0.13,0.134026574234547);
\draw[fill=color0,draw opacity=0] (axis cs:0.135,0) rectangle (axis cs:0.15,0.149763033175355);
\draw[fill=color0,draw opacity=0] (axis cs:0.155,0) rectangle (axis cs:0.17,0.165850673194614);
\draw[fill=color0,draw opacity=0] (axis cs:0.175,0) rectangle (axis cs:0.19,0.188715420432758);
\draw[fill=color0,draw opacity=0] (axis cs:0.195,0) rectangle (axis cs:0.21,0.211046990931575);
\draw[fill=color0,draw opacity=0] (axis cs:0.215,0) rectangle (axis cs:0.23,0.22755158951478);
\draw[fill=color0,draw opacity=0] (axis cs:0.235,0) rectangle (axis cs:0.25,0.250622820129547);
\draw[fill=color0,draw opacity=0] (axis cs:0.275,0) rectangle (axis cs:0.29,0.288848263254113);
\draw[fill=color0,draw opacity=0] (axis cs:0.295,0) rectangle (axis cs:0.31,0.314814814814815);
\draw[fill=color0,draw opacity=0] (axis cs:0.315,0) rectangle (axis cs:0.33,0.333333333333333);
\draw[fill=color0,draw opacity=0] (axis cs:0.335,0) rectangle (axis cs:0.35,0.354251012145749);
\draw[fill=color0,draw opacity=0] (axis cs:0.355,0) rectangle (axis cs:0.37,0.369950389794472);
\draw[fill=color0,draw opacity=0] (axis cs:0.375,0) rectangle (axis cs:0.39,0.391581632653061);
\draw[fill=color0,draw opacity=0] (axis cs:0.395,0) rectangle (axis cs:0.41,0.408831908831909);
\draw[fill=color0,draw opacity=0] (axis cs:0.415,0) rectangle (axis cs:0.43,0.434027777777778);
\draw[fill=color0,draw opacity=0] (axis cs:0.435,0) rectangle (axis cs:0.45,0.442307692307692);
\draw[fill=color0,draw opacity=0] (axis cs:0.455,0) rectangle (axis cs:0.47,0.469450101832994);
\draw[fill=color0,draw opacity=0] (axis cs:0.475,0) rectangle (axis cs:0.49,0.495024875621891);
\draw[fill=color0,draw opacity=0] (axis cs:0.495,0) rectangle (axis cs:0.51,0.502222222222222);
\draw[fill=color0,draw opacity=0] (axis cs:0.555,0) rectangle (axis cs:0.57,0.567353407290016);
\draw[fill=color0,draw opacity=0] (axis cs:0.595,0) rectangle (axis cs:0.61,0.611929307805597);
\draw[fill=color0,draw opacity=0] (axis cs:0.615,0) rectangle (axis cs:0.63,0.623268698060942);
\draw[fill=color0,draw opacity=0] (axis cs:0.635,0) rectangle (axis cs:0.65,0.643518518518518);
\draw[fill=color0,draw opacity=0] (axis cs:0.655,0) rectangle (axis cs:0.67,0.675977653631285);
\draw[fill=color0,draw opacity=0] (axis cs:0.675,0) rectangle (axis cs:0.69,0.697297297297297);
\draw[fill=color0,draw opacity=0] (axis cs:0.715,0) rectangle (axis cs:0.73,0.721244925575101);
\draw[fill=color0,draw opacity=0] (axis cs:0.735,0) rectangle (axis cs:0.75,0.748663101604278);
\draw[fill=color0,draw opacity=0] (axis cs:0.755,0) rectangle (axis cs:0.77,0.765217391304348);
\draw[fill=color0,draw opacity=0] (axis cs:0.775,0) rectangle (axis cs:0.79,0.797005988023952);
\draw[fill=color0,draw opacity=0] (axis cs:0.795,0) rectangle (axis cs:0.81,0.818320610687023);
\draw[fill=color0,draw opacity=0] (axis cs:0.815,0) rectangle (axis cs:0.83,0.836472602739726);
\draw[fill=color0,draw opacity=0] (axis cs:0.835,0) rectangle (axis cs:0.85,0.854970760233918);
\draw[fill=color0,draw opacity=0] (axis cs:0.875,0) rectangle (axis cs:0.89,0.888979591836735);
\draw[fill=color0,draw opacity=0] (axis cs:0.895,0) rectangle (axis cs:0.91,0.914505824877865);
\draw[fill=color0,draw opacity=0] (axis cs:0.915,0) rectangle (axis cs:0.93,0.931207632437861);
\draw[fill=color0,draw opacity=0] (axis cs:0.935,0) rectangle (axis cs:0.95,0.950620251947303);
\draw[fill=color0,draw opacity=0] (axis cs:0.955,0) rectangle (axis cs:0.97,0.974191433566434);
\draw[fill=color0,draw opacity=0] (axis cs:0.975,0) rectangle (axis cs:0.99,0.992896810903617);
\addplot [thick, color0, dashed]
table {%
0 0
1 1
};
\end{axis}

\end{tikzpicture}
    }
    \caption{Empirical calibration plot~\cite{ramos-reliableSupport-ForensicScience-2013} on the pre-trained classifier's scores on V2D. Applying PAVA~\cite{pav_brummer} results in oracle calibrated scores (b).}
    \label{fig:empcalplot}
\end{figure}
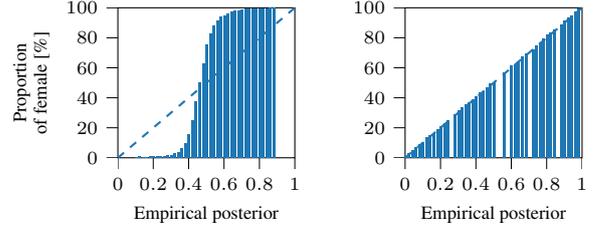\\
\textbf{The encoder} encodes an input x-vector $x$ into an embedding~$z$. In addition to compressing the input information, it tries to cheat the adversarial sex classifier.\\
\textbf{The decoder} reconstructs the original x-vector $x$ from the variable $z$ and a condition $w$. During the training $w=\tilde{y}$.\\
\textbf{The adversarial sex classifier} tries to predict, from the encoded vector~$z$, the sex class of the corresponding speech segment.
\begin{figure}[ht]
    \centering
    \includegraphics[width=8cm]{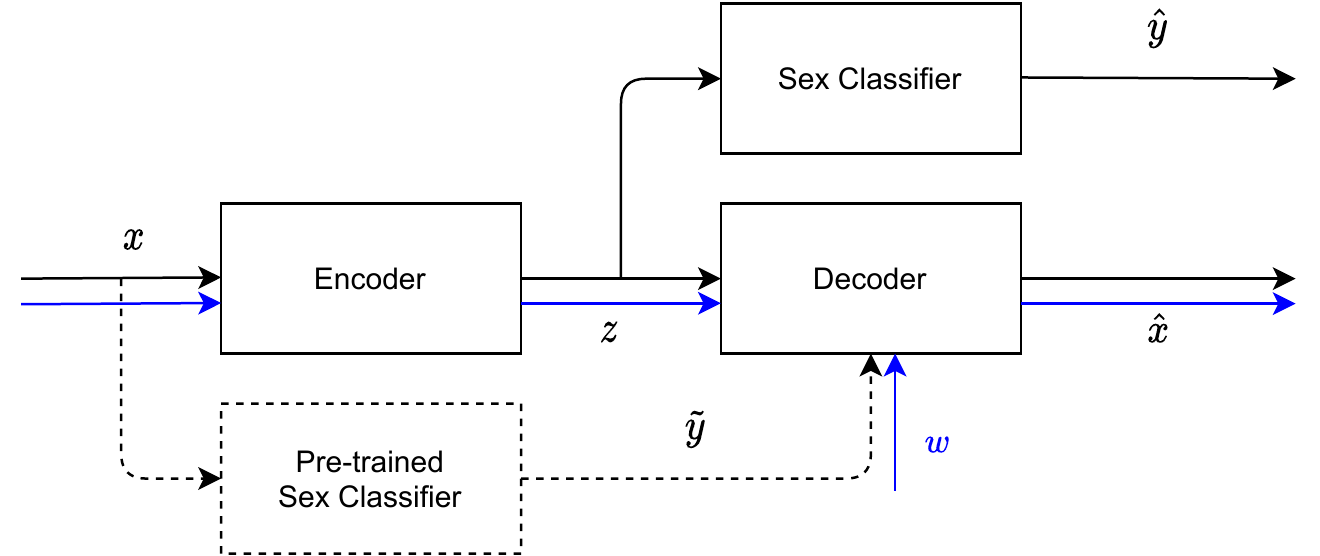}
    \caption{Illustration of the Adversarial Disentangling Autoencoder. The black arrows illustrate the forward flow during the training phase. The dashed lines represent the part that is not updated during training: the pre-trained sex classifier is used to extract posteriors $\tilde{y}$ for feeding the decoder. The blue arrows illustrate the forward flow during the protection phase. Sex attribute control is done by setting $w$.}
    \label{fig:net_diagram}
\end{figure}

The training and protection flows in the model are illustrated in Figure~\ref{fig:net_diagram}.
The sex information is disentangled from the rest using an adversarial training that opposes the encoder-decoder to the adversarial sex classifier. Thus, a first optimiser updates the neural parameters $\theta_{c}$ of the adversarial sex classifier with respect to the correct expected class prediction of $x$. Then, a second optimiser updates the parameters $\theta_{E}$ of the encoder and the parameters $\theta_{D}$ of the decoder in order to jointly cheat the adversarial sex classifier and reconstruct the x-vector. Therefore, the encoder-decoder and the sex classifier are trained in an adversarial manner in order to make the encoded vector $z$ sex-independent. The two objective functions are respectively:

\begin{equation}
    L_{d}(\theta_{c}|\theta_{E})=-\frac{1}{m}\sum_{i = 1}^{m   }\log(\hat{y}_{i}),
\end{equation}
\begin{equation}
    L(\theta_{E},\theta_{D}|\theta_{c})=\frac{1}{m}\sum_{i=1}^{m}\left(r(\hat{x}_{i},x_{i})-\log(1-\hat{y}_{i})\right),
\end{equation}

where $z = E_{\theta_{E}}(x)$ and $\hat{x} = D_{\theta_{D}}(z,\tilde{y})$ are respectively the output of the encoder and the decoder, 
$\hat{y}$ is the adversarial sex classifier predicted score and $r(\hat{x},x) = 1-\frac{<\hat{x},x>}{||\hat{x}||||x||}$ is the reconstruction error. During our experiments, both optimisers follow the stochastic gradient descent with the model parameters $\theta_{E}^{(t)}$, $\theta_{D}^{(t)}$ and $\theta_{c}^{(t)}$ being updated at each time step $t$ as follow:

\begin{align}
    \theta_{c}^{(t+1)}&=\theta_{c}^{(t)} \hspace{-0.35cm}&-& \eta \nabla_{\theta_{c}} L_{d}(\theta_{c}^{(t)}|\theta_{E}^{(t)},x^{(t)},y^{(t)}),\\
    \begin{bmatrix}
        \theta_{E}^{(t+1)} \\
        \theta_{D}^{(t+1)}
    \end{bmatrix}
    \hspace{-0.1cm}&=\hspace{-0.1cm}
    \begin{bmatrix}
        \theta_{E}^{(t)} \\
        \theta_{D}^{(t)}
    \end{bmatrix}
    \hspace{-0.35cm}&-& \eta\nabla_{\theta_{E},\theta_{D}} L(\theta_{E}^{(t)},\theta_{D}^{(t)}|\theta_{c}^{(t)},x^{(t)},y^{(t)}),
\end{align}
with $x^{(t)}$ the training sample and $y^{(t)}$ its corresponding class at time $t$.

\subsection{Attribute protection
}
\label{sec:protection}
To achieve zero-evidence \cite{Nautsch2020}, zero log-likelihood ratios (LLRs) need to result: to protect the sex attribute, when x-vectors are taken as evidence, the likelihood for an x-vector given the female proposition needs to equal the likelihood for 
the male proposition. Such zero LLRs imply that updating any prior belief with them results in posterior belief of the same value. In this way, using a sex classifier becomes useless. This idea is motivated from perfect secrecy \cite{shannon} and method validation in forensic sciences \cite{Willis-Guideline-Evaluative-Reporting-ENFSI-2015}.\\
In our case, the value of the decoder input $w$ permits to control in the x-vector the sex information which is expressed as a posterior $P(F|x)$. Different perspectives arise for setting $w$: when setting $w=\tilde{y}$, privacy is not preserved; for zero LLRs, the attacker posterior must remain the attacker prior $\pi$. 
One reflex could be to set $w$ to the attacker prior $\pi$, which however is assumed unknown. Therefore, in this work, we set for every x-vector $w=0.5$, a non-informative/objective prior. Yet, while discriminative sex information is reduced, zero-evidence is not necessarily ensured when the attacker treats the given posteriors as scores\footnote{Actually, if an attacker trains a calibration, zero-LLRs are implied if all scores are constant/equal.}.



\section{Experimental and Evaluation Setup}
\label{sec:global_exps}

This section first details the architecture, the training procedure of the proposed model and the corpora used (Section~\ref{sec:archi_train_corpora}). Then, evaluation protocols are described (Section~\ref{sec:eval_protocol}).

\subsection{Model architecture, training and corpora}
\label{sec:archi_train_corpora}
First, a standardisation and a length normalisation~\cite{danielgarciaromero2011lt} are applied on the x-vectors before being fed into the model. The encoder consists of a single dense layer with ReLU activation functions~\cite{nair2010rectified}. Batch-normalisation~\cite{pmlr-v37-ioffe15} is applied accross the resulting $128$ dimensional representation $z$. Then, the decoder takes as inputs the encoded vector $z$ concatenated with the sex posterior $\tilde{y}$, and consists of a single dense layer with a hyperbolic tangent for the activation function followed by a last length normalization. The adversarial classifier is composed with two dense layers. The first one has $64$ units with ReLU activation functions, and the second one has $1$ unit with a sigmoid activation function. The model is trained using two standard stochastic gradient descents. Learning rates are set to $10^{-4}$ and the momentums are $0.9$. A subset V1D (61616 segments per class) of VoxCeleb~\cite{voxceleb} is used to train the external sex classifier\footnote{A one layer perceptron with a sigmoid as activation function. Scores are oracle calibrated using the pool-adjacent-violator algorithm~\cite{pav_brummer}.} that extracts the posteriors $\tilde{y}$. A subset V2D (397032 segments per class) of VoxCeleb2 development part~\cite{voxceleb2} is used to train the adversarial disentangling autoencoder model. A subset V2T (9120 female segments and 22559 male segments) of VoxCeleb2 testing part is used to test how well the model protects the attribute.

\subsection{Evaluation protocol}
\label{sec:eval_protocol}
As explained in Section~\ref{sec:protection}, the decoder input $w$ is set to 0.5 
in order to hide the sex information in the reconstructed x-vector. The following explains how to assess the level of protection and its impact on the utility~i.e. the ASV performance.\\

\noindent\textbf{Attribute privacy preservation assessment:\\}
In order to assess the protection ability, several metrics are computed from the sex classifier scores. 
We compute the area under the receiver operating characteristic curve (AUC). An AUC of $50\%$ would confirm the random prediction of the classifier.
The $\Cllrmin$ is also computed as a discrimination cost~\cite{brummer2006application}. Recently, the Zero Evidence Biometric Recognition Assessment framework was introduced~\cite{Nautsch2020}. Originally presented in the context of speaker identity preservation, it can be applied to any binary decision task. It provides a measure $\Dece$ of the expected amount of private information disclosed as well as the worst-case score $l_w$ i.e. the strongest strength-of-evidence in a set of segments. The latter comes with a categorical tag for better interpretation. For more details on these measures, refer to~\cite{Nautsch2020}.\\ 
The Mutual Information (MI) can also be used as a privacy measure~\cite{8007053}. In our experiments, MI is used to measure the amount of information that the x-vector is sharing with the sex variable~$y$. The MI between the x-vectors' components and $y$ is estimated~\cite{ross2014mutual,scikit-learn} and its average over the x-vector's dimensions is given.\\
\noindent\textbf{Automatic speaker verification evaluation:}\\
As a first example of an attribute-driven privacy preservation application, we focus on ASV. An important aim of this work is indeed to hide the sex information in the x-vector representations while maintaining automatic speaker verification ability. Thus, ASV performance with the protected x-vectors are measured. The commonly used Probabilistic Linear Discriminant Analysis (PLDA)~\cite{pldaIoffe} is used to compute the likelihood-ratios from the comparisons of enrolment and probe x-vectors.
\begin{table}[ht]
\caption{Description of the ASV datasets.}
\label{tab:statasvsets}
    \begin{tabular}{ccccc}
        \cline{2-5}
        & \multicolumn{2}{c}{Number of segments} & \multicolumn{2}{c}{Number of speakers} \\
        \cline{2-5}
         & Enrolment & Test & Enrolment & Test \\
        \hline 
        Male & 11282 & 11277 & 81 & 81 \\
        Female & 4558 & 4562  & 39 & 39 \\
        \hline
    \end{tabular}
\end{table}

Both enrolment and probe x-vectors are converted. The Equal Error Rate (EER) and the $\Cllrmin$ are measured to assess the ASV
. The PLDA have been trained on $200,000$ x-vectors randomly chosen from the VoxCeleb 1 and 2 training subsets. Their dimension is beforehand reduced to $128$ with a linear discriminant analysis. Details on the enrolment and test sets are shown in Table~\ref{tab:statasvsets}.

\section{Results}
\label{sec:results}

This section presents the results of the sex attribute concealment (Section~\ref{sec:dis_eval}) and the ASV performance on the protected x-vectors (Section~\ref{sec:asv_eval}).

\begin{table*}[h]
    \centering
    \caption{Effects of the reconstruction and the sex protection on the ability to predict, using a pre-trained classifier, the sex from x-vector on V2D and V2T. The first row reports results on original x-vectors. The second row refers to the x-vectors that have been passed through the system but without transformation ($w=\tilde{y}$). The last row refers to the protected x-vectors ($w=0.5$).}
    \label{tab:classif_results}
    \scalebox{1}{
    \begin{tabular}{ccccccc}
    
            \cline{2-7}
            & \multicolumn{2}{c}{\makecell{AUC {\tiny $10^{-2}$}}}
            & \multicolumn{2}{c}{\makecell{$\Cllrmin${\tiny $10^{-2}$}}}
            & \multicolumn{2}{c}{\makecell{($\Dece$, $\log_{10}(l_w)$, tag)}}
            \\
            \cline{2-7}

             & V2D & V2T & V2D & V2T & V2D & V2T\\
            \hline
            Original vector $x$ & 98.84 & 99.09 & 15.97 & 13.18 & (0.596, 2.910, C) & (0.619, 3.538, C)  \\
            
            $w = \tilde{y}$ & 98.00 & 97.29 & 19.19 & 17.60 & (0.570, 2.859, C) & (0.584, 3.554, C) \\
    
            $w=0.5$ & 49.19 & 55.23 & 99.79 & 98.64 & (0.001, 0.813, A) & (0.009, 0.393, A) \\
            
            \hline
    \end{tabular}
    }
\end{table*}
\begin{figure}[h]
\centering
\subfloat[Original data]{
    \begin{tikzpicture}[font=\scriptsize]
            \begin{axis}[width=0.22\textwidth, height=0.2\textwidth,ybar, ymin=0, ymax=7500, xmin=-0.01, xmax=1.01, xlabel={Posterior},ylabel={frequency},xtick pos=left,ytick pos=left]
            \addplot +[draw=none, red, bar width=0.02, opacity=0.5] table {hist/hist0O.txt};
            \end{axis}
            \begin{axis}[axis line style={draw=none}, width=0.22\textwidth, height=0.2\textwidth, ybar, ymin=0, ymax=7500, xmin=-0.01, xmax=1.01,      yticklabels={,,}, xticklabels={,,}, ytick style={draw=none}, xtick style={draw=none}]
            \addplot+[draw=none, blue, bar width=0.02, opacity=0.5] table {hist/hist1O.txt};
            \end{axis}
\end{tikzpicture}
}
\hfill
\subfloat[With protection ($w=0.5$)]{
    \begin{tikzpicture}[font=\scriptsize]
            \begin{axis}[width=0.22\textwidth, height=0.2\textwidth ,ybar, ymin=0, ymax=7500, xmin=-0.01, xmax=1.01, xlabel={Posterior}, ylabel={frequency},xtick pos=left,ytick pos=left]
            \addplot +[draw=none, red, bar width=0.02, opacity=0.5] table {hist/hist0Z.txt};
            \end{axis}
            \begin{axis}[axis line style={draw=none}, width=0.22\textwidth, height=0.21\textwidth, ybar, ymin=0, ymax=7500, xmin=-0.01, xmax=1.01,      yticklabels={,,}, xticklabels={,,}, ytick style={draw=none}, xtick style={draw=none}]
            \addplot+[draw=none, blue, bar width=0.02, opacity=0.5] table {hist/hist1Z.txt};
            \end{axis}
\end{tikzpicture}
}
\caption{Histograms of the calibrated scores from the sex classifier on V2T original (a) and protected (b)
. Female scores are in blue and male scores are in red. When protection is applied (b), the score distributions overlap almost perfectly.}
\label{fig:distrib}
\end{figure}
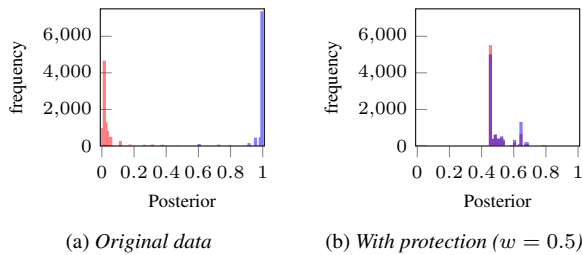

\subsection{Attribute privacy preservation results}
\label{sec:dis_eval}


Table~\ref{tab:classif_results} reports the sex classifier's AUC, the $\Cllrmin$ and the ZEBRA measures obtained with original, unprotected reconstructed (i.e. the x-vectors have been passed through the system but with $w=\tilde{y}$) and protected x-vectors (i.e. with $w=0.5$). 
At first, the AUC (lower than 0.5 for V2D with $w=0.5$), suggests that the class labels might be swapped/the direction of LLR scores is negated. The association of class labels to particular posterior values is arbitrarily fixed but since the method can violate\deleted[comment=grammar]{s} this \emph{fixed} assumption, inference using these assumptions can suffer\deleted[comment=grammar]{s}. We thus report results from labels' association that results in lower $\Cllrmin$ and bigger $\Dece$.\\
With $w=\tilde{y}$, the sex from whom the reconstructed x-vector comes from can still be detected correctly. With $w=0.5$, the AUC is getting close to 50\% which corresponds to random prediction. There is also a significant increase of the $\Cllrmin$ which confirms the difficulty to separate the score distributions also illustrated in Figure~\ref{fig:distrib}. Moreover, the drop of expected privacy disclosure ($\Dece$) corroborates the difficulty for the adversary to benefit from the scores and the low strongest strength-of-evidence (A) suggests that no x-vectors are left with a poor protection.
\begin{table}[h]
        \centering
        \caption{Mutual information measures between the x-vectors and the sex class variable $y$ on V2D and V2T.}
    \label{tab:mi}
        \begin{tabular}{ccc}
            \cline{2-3}
            & \multicolumn{2}{c}{$I(\hat{x},y)$ {\tiny $10^{-2}$} [bit per dimension]}\\
            \hline
             &  V2D  & V2T \\
            \hline
            Original vector $x$ & 18.7 & 19.0 \\
            $w = \tilde{y}$ & 20.3 & 19.81  \\
            $w = 0.5$ & 1.0 &  1.90 \\
            \hline
        \end{tabular}
\end{table}\vspace{-0.1cm}
The mutual information measures are shown in Table~\ref{tab:mi}. There is a significant decrease of mutual information when $w$ is set to 0.5. Therefore, the system seems to reduce the dependency between the protected x-vector and the sex variable resulting in a more \emph{sex-independent} speaker representation.
\subsection{Automatic speaker \replaced[comment=typping error]{v}{s}erification results}
\label{sec:asv_eval}
Table~\ref{tab:asvresults} shows the results obtained for the ASV task on the transformed x-vectors. For both unprotected reconstructed ($w=\tilde{y}$) x-vectors and protected ($w=0.5$) x-vectors the ASV performance slightly decreases. With protection, the EER increases from 1.72 to 2.36 and the $\Cllrmin$ increases from 0.067 to 0.097. The ASV performance is slightly better for $w=0.5$ in comparison to $w=\tilde{y}$. This suggests that the lose of performance 
is mostly due to the reconstruction error and that segments comparison for speaker verification could go without considering most of the sex information.
\begin{table}[H]
    \centering
    \caption{Effects of the reconstruction and the sex concealment on the ASV.}
    \label{tab:asvresults}
    \begin{tabular}{ccc}
        \hline
         & EER [\%] & $\Cllrmin$ \\
        \hline
        Original vector $x$ & 1.72 & 0.067 \\
        $w = \tilde{y}$ & 2.89 & 0.118\\
        \textbf{$w = 0.5$} & 2.36 & 0.097 \\
        \hline
    \end{tabular}
\end{table}

\section{Conclusion}
\label{sec:conclusion}
In this paper we have introduced the \emph{attribute-driven privacy preservation} as the idea of enabling a speaker to hide only a few personal aspects in its voice representation while maintaining the remaining particularities and the performance of a desired task. As a first solution, we presented an adversarial autoencoding approach that disentangles and hide a given binary attribute 
in a x-vector neural embedding. This method is based on an encoder-decoder architecture combined with an additional classifier that tries to predict the attribute class from the encoded representation. Both are trained in an adversarial manner to make the encoded representation \emph{attribute-independent}. This approach\replaced[comment=grammar]{ has}{ have} been experimented on the sex attribute and with the aim of enabling speaker verification. Through the experiments conduced on the Voxceleb dataset, it has been shown that setting the sex variable representing the posterior to 0.5 for all segments enables to diminish the sex information in the speaker representation with only a slight alteration of the automatic speaker verification performance.\\
Even though the proposed approach reduces the amount of sex information contained in a x-vector, it is not guaranteed that an attacker in possession of the transformed data and their original sex class can not take advantage of the remaining sex information and train a new classifier that will detect the original sex. 
In future works, this approach will be tested on other attributes such as the age or the regional accent. The method will also be combined to a speech synthesizer, therefore, the attribute-driven privacy preservation will output speech signal just like speech anonymisation~\cite{Tomashenko2020}.

\section{Acknowledgements}

This work was supported by the VoicePersonae ANR-18-JSTS-0001 and the Robovox ANR-18-CE33-0014 projects.

\bibliographystyle{IEEEtran}
\balance\bibliography{refs_pg_noe}

\end{document}